\documentclass[aps,pra, twocolumn, nofootinbib, showpacs]{revtex4-1} 
\usepackage[unicode=true,pdfusetitle,bookmarks=true,bookmarksnumbered=false,bookmarksopen=false,breaklinks=false,pdfborder={0 0 0},backref=false,colorlinks=false] {hyperref}
\hypersetup{colorlinks,linkcolor=myurlcolor,citecolor=myurlcolor,urlcolor=myurlcolor}
\usepackage{ulem,braket,colortbl,amsthm,amsmath,cleveref,amssymb,txfonts}\definecolor{myurlcolor}{rgb}{0,0,0.7}
\usepackage{bbm}
\usepackage{graphics,graphicx}
\usepackage{color}
\usepackage{amssymb}
\usepackage{amsthm}
\usepackage{amsfonts}
\usepackage{float}
\usepackage{xcolor}
\usepackage{graphicx}
\usepackage[format = plain,labelfont = bf,up, textfont = normal , up, justification=raggedright, singlelinecheck =false]{caption}
\usepackage{subcaption}
\usepackage{tabularx}
\usepackage{amsmath}
\usepackage{braket}

\usepackage{graphicx}
\usepackage[utf8x]{inputenc}
\usepackage{color,soul}
\usepackage{amsmath}
\usepackage{braket}
\usepackage{latexsym}
\usepackage{amssymb}
\usepackage{amsthm}
\usepackage{xcolor}
\usepackage{bm}
\usepackage{graphics,epstopdf}
\usepackage{color}\usepackage{amsmath}
\usepackage{enumitem}
\usepackage{fmtcount}
\usepackage{booktabs}
\usepackage{epsfig}

\theoremstyle{plain}

\begin{document}
\title{Quantum phase estimation in presence of glassy disorder}

\author{Soubhadra Maiti$^{1,2}$, Kornikar Sen$^1$, Ujjwal Sen$^1$}
\affiliation{$^1$Harish-Chandra Research Institute, A CI of Homi Bhabha National Institute, Chhatnag Road, Jhunsi, Allahabad 211 019, India\\
 $^2$Indian Institute of Science Education and Research, Pune 411 008, India}

\begin{abstract}

We investigate the response to noise, in the form of glassy disorder present in circuit elements, in the success probability of the quantum phase estimation algorithm, a subroutine used to determine the eigenvalue - a phase - corresponding to an eigenvector of a unitary gate.
We prove that when a large number of auxiliary qubits are involved in the circuit, the probability does not depend on the actual type of disorder but only on the mean and strength of the disorder. For further analysis, we consider three types of disorder distributions: Haar-uniform with a circular  cut-off, Haar-uniform with an elliptical or squeezed
cut-off, and spherical normal.
There is generally a depreciation of the disorder-averaged success probability in response to the disorder incorporation. 
Even in the presence of the disorder, increasing the number of auxiliary qubits  helps  to get a better precision of the phase, albeit to a lesser extent (probability) than that in the clean case.
%
We find a concave to convex transition in the dependence of probability on the strength of disorder,
and a log-log dependence is witnessed between the point of inflection and the number of auxiliary qubits used. 

\end{abstract}

\maketitle



\section{Introduction}
\label{sec1}

There are various aspects of  quantum mechanics that make it strikingly distinct from its classical cousin. Harnessing the power of these quantum characteristics provide or are believed to provide significantly improved efficiencies  in numerous quantum algorithms with respect to the corresponding classical algorithms~\cite{ref1,ref20,watrous}. Examples include  the Deutsch–Jozsa algorithm~\cite{deutsch-algo1,deutsch-algo2}, Shor's algorithm~\cite{ref4}, Grover's algorithm~\cite{grover-algo}, etc. 


The quantum phase estimation algorithm (QPEA)~\cite{ref2,ref6,R2} is tailored 
to determine the eigenvalues, which are phases, of a given unitary operator. 
%
%
Quantum phase estimation 
is applicable to problems that can effectively be turned into solving an eigenvalue equation
such as computing  molecular spectra~\cite{exp5,exp6,exp7}.
Shor's algorithm for prime factorization of integers, which classically is believed to take a number of steps that is exponential in the input size, also utilizes the QPE subroutine for its 
speed-up~\cite{ref1,ref4}. QPEA has applications in quantum sampling \cite{qs1,qs2}.  Furthermore, the QPE problem is relevant to clock synchronization, magnetometry, etc.~\cite{ref11,ref12,ref13,ref14,ref15,ref16,ref17,ref18,ref19}. Experimental implementations of the QPEA include those in Refs.~\cite{exp1,exp2,exp3,exp4,exp5,exp6,exp7}.



In realistic situations, the computational machines do not operate as we would like them to do ideally. The reasons behind the flaws might be some imperfections present in the building blocks or unrecognized or uncontrolled interactions of the devices with the environment. 
It is thus reasonable to explore the effects of noise on the performance of the quantum phase estimation process.
%
QPE has been analyzed 
by considering various 
noise models \cite{ref23,ref24,ref30,ref33,ref34,ref36,ref37,ref38,ref39,ref28,ref28B,ref27,ref29,ref31}. 
For example, the effects of white, bit flip, phase flip, and bit-phase flip noise on QPEA are discussed in \cite{ref39,ref37}. In optical interferometry, the fundamental limits to  precision of phase estimation for light with definite photon number
is discussed in~\cite{optimalQPE}, in the presence of photon loss. In  absence of entanglement in the system,  error in the estimated phase scales as the inverse square root of the photon number;
while, taking advantage of entanglement, it can be improved to an inverse photon number scaling, 
known as the Heisenberg limit~\cite{HeisenbergLimit}.  
Using the Cram{\'e}r-Rao bound~\cite{Fisher1} and the concept of Fisher information~\cite{Fisher2,Fisher3}, several studies have demonstrated that the expected precision can deviate from the Heisenberg limit in  presence of noise~\cite{ref28,ref28B,ref30}. 

In this paper, we study the effects of noise induced by 
``glassy disorder", in the circuit components, on the quantum phase estimation algorithm. 
We assume that only the Hadamard gates are affected by such disorder.
Three types of glassy disorders are considered: uniform (Haar-uniform with a circular cut-off), squeezed (Haar-uniform with an elliptical cut-off), and spherical normal (von Mises-Fisher).
We find that the depreciation in  probability, with increasing disorder degree,  of correctly predicting the phase is related to the number of auxiliary qubits used in the algorithm.
{We show that for a large number of auxiliary qubits, the disorder-averaged probability only depends on the strength of the disorder and not on the particular form of the disorder.}
In all cases of disorder, when the noise is weak, the probability changes slowly with the degree of disorder. However, after crossing a threshold value of the strength of disorder, a sudden change in probability is observed: the rate of fall of  probability with increasing disorder, gets faster. Further increasing the disorder, a cut-off is encountered, after which the probability becomes almost constant. These thresholds and cut-off values depend on the number of auxiliary qubits. In particular, the number of participating auxiliary qubits has a log-log dependence on the disorder amount at which the concave to convex transition is noticed. The minimum probability of successful detection of phase, with a certain precision, exhibits the same feature. 
We see that larger the number of auxiliary qubits used, greater is the effect of disorder on the probabilities. 
{We also observe that the squeezed disorder has more effect on the algorithm than the non-squeezed uniform disorder, for the same projected areas of the two disorders.
}
{The possibility of experimental differentiation between the types of disorder active in the circuit will depend on the available precision of instruments, but higher precision - possibly prohibitively high - will be needed with increase in the number of auxiliary qubits. It will also depend on the range of the strength of the disorder that is active. }

{The rest of the paper is organized as follows. In Sec.~\ref{sec2}, we briefly review the quantum phase estimation algorithm. Sec.~\ref{sec3} consists of a description of the disorder that we impose on the algorithm. In Sec.~\ref{sec4}, we prove that whatever be the disorder, the response will be the same, provided the strength of disorder is fixed and the number of auxiliary qubits used is large.  In Sec.~\ref{sec5}, we focus our attention to a fixed set of disorder distributions, and provide  short discussions on three specific disorder distributions. In Sec.~\ref{sec6}, we discuss the response of the algorithm to insertion of disorder from the particular  distributions. A conclusion is presented in Section~\ref{sec7}. Three appendices discuss the supplementary material.}


\section{Phase estimation algorithm}
\label{sec2}
In this section, we will briefly recapitulate the quantum Fourier transform-based QPEA~\cite{ref6}.
Suppose $U$ is a unitary operator acting on an $n$-qubit Hilbert-space. One eigenvector of $U$ is $\ket{\psi}$, and the corresponding eigenvalue is $e^{2\pi ip}$, where $p \in [0,1)$.
Thus we have
\begin{equation*}
    U\ket{\psi} = e^{2\pi ip}\ket{\psi}.
\end{equation*}
The aim of the algorithm is to estimate the phase, $p$, of the eigenvector, $\ket{\psi}$. 
A controlled-$U$ gate,  $\Lambda_m(U)$, is needed in the algorithm. Auxiliary qubits are also needed, whose number depends on the precision required, and in turn the available number of auxiliary qubits that can be controlled decides the precision. Let the number of available auxiliary qubits be $m$. $\Lambda_m(U)$ acts on the composite system consisting of $m+n$ qubits, where the first $m$ are the auxiliary control qubits and the remaining $n$ are the target qubits. We denote the basis of each of the single-qubit Hilbert spaces as $\{\ket{0},\ket{1}\}$.
 The operation of $\Lambda_m(U)$ is given by
\begin{equation}
\label{je-katha-tomake-bolbo}
    \Lambda_m(U) \ket{k}\ket{\psi} = \ket{k}\left(U^k \ket{\psi}\right) = e^{2\pi ik p}\ket{k}\ket{\psi}.
\end{equation}
Here, $k$, whenever used outside the ket \(|k\rangle\), is the decimal number corresponding to the binary number representing the $m$-qubit auxiliary state \(|k\rangle\). Thus, $k \in \{ 0, 1, ... , 2^m - 1 \}$. 

To perform the algorithm, the state $\ket{\chi_0}=\ket{0}^{\otimes m}\ket{\psi}$ is prepared initially. Next, Hadamard gates are acted upon the auxiliary qubits, transforming the initial state, $\ket{\chi_0}$, to $\ket{\chi_1}=\frac{1}{2^{m/2}}\sum_{k=0}^{2^m -1}\ket{k}\ket{\psi}$. The information about $p$ is encoded in the $m$ auxiliary qubits via a single application of the controlled-$U$ gate. 
The controlled unitary transformation, $\Lambda_m(U)$, results in a phase kickback effect:
\begin{align}\label{lambdaU_state}
   \ket{\chi_2} \equiv \Lambda_m(U)\ket{\chi_1} 
    = \frac{1}{2^{m/2}}\sum_{k=0}^{2^m -1}e^{2\pi ikp}\ket{k}\ket{\psi}.
\end{align}
At this point, it is possible to discard the state $\ket{\psi}$. $\ket{\psi}$ can be reused later if needed (and if possible). 

The quantum Fourier transformation is defined as follows:
\begin{equation*}
    \text{QFT}_{2^m}\ket{j}=\frac{1}{2^{m/2}}\sum_{k=0}^{2^m-1}e^{2\pi i jk/2^m}\ket{k}.
\end{equation*}
Applying $\text{QFT}_{2^m}^{\dagger}$ on the auxiliary qubits, 
one gets
\begin{equation*}
    \ket{\zeta}=\sum_{j=0}^{2^m-1}\left( \frac{1}{2^m}\sum_{k=0}^{2^m-1}e^{2\pi ik(p - j/2^m)}\right)\ket{j}.
\end{equation*} 
Finally, a projective measurement can be performed on the auxiliary states of $\ket{\zeta}$, in the computational basis, 
$\{\ket{j}\}_{j=0}^{2^m-1}$. The outcome corresponding to the output state $\ket{j}$, which is expected to represent the estimated phase, can be defined to be $\frac{j}{2^m}$. Then the probability of getting the outcome $\frac{j}{2^m}$ is
\begin{equation}\label{noiseless-prob}
    p_j = \left|{\frac{1}{2^m}\sum_{k=0}^{2^m-1}e^{2\pi ik(p-j/2^m)}}\right|^2.
\end{equation}
For the special case when $p = j/2^m$, $p_j$ is exactly equal to $1$, and hence in that case the outcome can accurately estimate the phase.
But in general, $p = j/2^m + \delta$, where $\delta$ is a real number such that $\left|{\delta}\right| \leq 2^{-(m+1)}$, and $j$ is an arbitrary integer satisfying $0\leq j\leq 2^m-1$. In that case, the probability of the outcome $j/2^m$ is $p_j \geq 4/\pi^2 > 0.4$. Thus the minimum probability of successfully gathering information about the phase of the eigenvector, $\ket{\psi}$, with accuracy up to $m$ binary points, is \(p_{min} = 4/\pi^2\), that is when $\delta=2^{-(m+1)}$ {[For the proof see Appendix D]}.
not of the form $j/2^m$,
increasing $m$ leads to better precision.
The following list shows the dependence of 
precision of the estimated phase 
on $m$: 

\begin{center}
\begin{tabular}{ |c|c|c|c|c|c|} 
\hline
$m$ & 13 & 18 & 21 & 24 & 27\\
\hline
precision & {$10^{-5}$} & {$10^{-6}$} & {$10^{-7}$} & {$10^{-8}$} & {$10^{-9}$}\\  
\hline
\end{tabular}
\end{center}

\section{Incorporation of disorder}
\label{sec3}

The central focus of this work is to analyze how imperfections disturb the QPEA.
The special class of disorders which we consider to be present in the circuit are referred to as
``glassy" 
disorder. It has also been termed in the literature as ``quenched'' disorder.
A system parameter is said to be glassy disordered when the equilibrium time of the disorder in the system is much larger than the typical observation time. This means that a particular realization of the disordered parameters do not change (appreciably) during the time of the observation. The values may change appreciably after a long time but that range of time is not in the domain of our interest.

To incorporate defects, we assume that the Hadamard gates acting on the initial $m$ auxiliary qubits' states, $\ket{0}^{\otimes m}$, are affected by noise, and as a result, instead of transforming the state $\ket{0}$ ($\ket{1}$) to $\ket{+}=\frac{\ket{0}+\ket{1}}{\sqrt{2}}$ $\left(\ket{-}=\frac{\ket{0}-\ket{1}}{\sqrt{2}}\right)$, the gate maps $\ket{0}$ ($\ket{1}$) to another state, $|\xi (\theta,\phi)\rangle$ ($|\xi^{\perp} (\theta,\phi)\rangle$), on the surface of the Bloch sphere. Here, $|\xi^{\perp} (\theta,\phi)\rangle$ is a vector (unique up to a phase) orthogonal to $|\xi (\theta,\phi)\rangle$, and $(\theta,\phi)$ are spherical polar coordinates on the Bloch sphere. 
We denote such ``noisy'' Hadamard gates as $H(\theta,\phi)$ and define the operation of $H(\theta,\phi)$ on the single qubit states, $\ket{0}$ and $\ket{1}$, in the following way:
\begin{eqnarray}
    H(\theta,\phi)\ket{0}=|\xi(\theta,\phi)\rangle := \cos\frac{\theta}{2}\ket{0} + e^{i\phi} \sin\frac{\theta}{2}\ket{1},\label{eq1}\\
    H(\theta,\phi)\ket{1}=|\xi^\perp(\theta,\phi)\rangle  := \sin\frac{\theta}{2}\ket{0} - e^{i\phi} \cos\frac{\theta}{2}\ket{1}.
\end{eqnarray}
The phase of \(|\xi^{\perp} (\theta,\phi)\rangle\) has been arbitrarily chosen and fixed, while defining the noisy Hadamard. The case when $\theta = \frac{\pi}{2}$ and $\phi = 0$, represents the mapping under the ``noiseless'' (i.e., the ordinary) Hadamard gate. In the disordered case, each pair of angles, $(\theta,\phi)$, defining the operation of the noisy Hadamard gates on the auxiliary qubits, will be chosen from an appropriate distribution depending on the type of the disorder distribution that is active. This can also be thought of as choosing the state $|\xi(\theta,\phi)\rangle$ or $|\xi^\perp(\theta,\phi)\rangle$ from the surface of the Bloch sphere.
Generally, the zenith angle, $\theta$, and the azimuthal angle, $\phi$, being the spherical polar coordinates, belong in the ranges $ \left[0, \pi\right]$ and \([ 0, 2\pi)\) respectively, but the ranges may get shortened in case of special disorders.


The quantity of interest, in our case, is the probability of obtaining correct output which can represent the phase accurately up to $m$ binary places. 
We denote each of the zenith and azimuthal angles as $\theta_i$ and $\phi_i$, which defines the operation of the noisy Hadamard gate, $H_i(\theta_i,\phi_i)$, on the \(i^{\text{th}}\) auxiliary qubit. Subsequently, after  measurement of the final state in the computational basis $\{\ket{j}\}_{j=0}^{2^m-1}$, the probability of obtaining the output state $\ket{j}$ is  
\begin{equation}\label{prob_j}
    p'_j(m) = \frac{1}{2^m}\prod_{i = 1}^{m}\left[ 1 + \sin\theta_i \cos\left[2^i \pi\left(p - \frac{j}{2^m}\right) + \phi_i\right] \right].
\end{equation}
For a  derivation, see Appendix~A.




For glassy disorder, the disorder average of a quantity is defined as the average of the quantity evaluated over the distribution of the disorder under discussion. Thus, the disorder-averaged probability can be obtained as
\begin{equation}
    q_j(m) = \idotsint p'_j(m) \prod_{i=1}^m f(\theta_i,\phi_i) \sin\theta_i d\theta_i d\phi_i, \label{eq4}
\end{equation}
where $f(\theta_i, \phi_i)$ is the probability density function of the distribution that the disordered system parameter follow. Keeping  realistic scenarios in mind, we assume that each pair of the angles, $(\theta_i,\phi_i)$, follow the same disorder distribution.

In the same manner, we can determine the minimum probability of obtaining the phase, accurate up to $m$ binary points, in presence of disorder, as
\begin{equation*}
    p'_{min}(m) = \frac{1}{2^m}\prod_{i = 1}^{m}\left[ 1 + \sin\theta_i \cos\left[ \frac{2^i \pi}{2^{m+1}} + \phi_i\right] \right].
\end{equation*}
The disorder-averaged value of $p'_{min}$ is denoted as $q_{min}$ and defined as
\begin{equation*}
    q_{min}(m) = \idotsint p'_{min}(m) \prod_{i=1}^m f(\theta_i,\phi_i) \sin\theta_i d\theta_i d\phi_i.
\end{equation*}

\section{When all disorders have equivalent response}
\label{sec4}
Here we want to investigate whether different types of disorder distributions affect the QPE algorithm differently. We will invoke the central limit theorem, which we briefly recapitulate below, for completeness.
\\


\noindent \textit{\textbf{Central limit theorem}:} Let $X_i~(i = 1,2,\ldots,n)$ be independent random variables with means, $\mu_i$, and finite standard deviations, $\sigma_i$. Then the sum $S_n = X_1 + X_2 + \ldots + X_n$ asymptotically follows the normal distribution with mean $n\mu= \sum_i \mu_i$ and standard deviation $\sqrt{n}\sigma = \left(\sum_i \sigma_i^2\right)^{1/2}$.
We can write the statement as 
\begin{equation}\label{CLT}
    S_n \sim \mathcal{N}(n\mu,\sqrt{n}\sigma),
\end{equation}
\noindent regardless of the population distribution, provided the sample size is large and that the sample standard deviations are finite.
\\

Reverting to the QPEA, the probability of getting a particular outcome, $j$, in presence of glassy disorder, is given in Eq.~\eqref{prob_j}. Natural logarithm on both sides of Eq. \eqref{prob_j} gives
\begin{equation*}
    \ln p'_j(m) = \sum_{i=1}^m x_i - m\ln 2,
\end{equation*}
where $x_i = \ln \left[1+\sin\theta_i\cos(2^i\pi\delta+\phi_i)\right]$. Since $\theta_i$s and $\phi_i$s are independently distributed, the variables $x_i$s are also independent of each other.
If the distribution of the random variables $\{x_i\}$ has mean $\mu^*$ and finite variance $\sigma^{*2}$, then according to the central limit theorem, we can write, for large \(m\),
\begin{equation}\label{log_p-normal}
    \ln p'_j(m) \sim \mathcal{N}\left(m(\mu^* - \ln 2), \sqrt{m}\sigma^{*}\right).
\end{equation}
Therefore, $p'_j(m)$ approximates a \textit{log-normal distribution}, 
as $m \to \infty$. Hence, as long as the variables $\theta_i$ and $\phi_i$ are independently chosen from a distribution so that the \(x_i\) have finite variances,
mean of $p'_j(m)$ approaches 
\begin{equation}
\label{banamali-tumi-parajaname-haiyo-radha}
\exp\left(m\left(\mu^* + \frac{\sigma^{*2}}{2} - \ln 2\right)\right),
\end{equation}
which only depends on the mean and variance of the distribution of the $\{x_i\}$ and is independent of other particulars of that 
distribution. This is true in our noise model since the mean of the distribution of the angles is taken to be at $\theta =\pi/2,\phi=0$, which corresponds to the state $\ket{+}$, and the standard deviation of the distribution of the \(x_i\) is also finite and fixed for a given setting of the apparatuses.
It is evident from Eq.~\eqref{eq4} that $q_j(m)$ is given by the expression in~(\ref{banamali-tumi-parajaname-haiyo-radha}), for large \(m\).
We therefore have the  following corollary.\\

\noindent \textit{\textbf{Corollary}:} Irrespective of the disorder distribution of the glassy disorder in the noisy Hadamard gates of the quantum phase estimation algorithm, the disorder-averaged success probability of finding the phase correct to a sufficiently high precision, depends only on the mean of the disorder distribution.\\



The precision of the phase - in turn - increases with an increase in  the number of auxiliary qubits used. But when a relatively small number of auxiliary qubits are considered, the effect of distinct disorder distributions on the QPEA success probability may have significant differences. Below, we will examine the effect of three classic disorder distributions followed by the angles $(\theta_i,\phi_i)$, for the case when \(m\) is not high.

\section{Different disorder distributions}
\label{sec5}


We consider three different distributions for the pair of angles $(\theta,\phi)$: Haar-uniform, squeezed, and von Mises-Fisher. Since the noiseless case is at the point for which 
$\theta=\pi/2$ and $\phi=0$, we fix the mean of all the disorder distributions to be at that point.
From now on, we use the notation $(1,\theta,\phi)$, to denote a point on the Bloch sphere that represents the state written in the RHS of Eq.~\eqref{eq1}.
The variance of any of the distributions is defined in the usual way, and remembering that the relevant distance between points on the Bloch sphere - for calculation of the variance of a disorder distribution - are along great circles on the unit sphere. 
The distances for finding the variance of a disorder distribution are calculated from the mean of the distribution, which is always chosen to be 
%
%
$(1,\pi/2,0)$, i.e., the state $\ket{+}$. (Therefore, instead of considering the noisy Hadamard gate directly, we use the state \(|\xi(\theta,\phi)\rangle\) for calculating the variance of the distribution: see Eq.~(\ref{eq1}).)
To make the mathematics for calculating the variance simpler,
we first consider the distributions to have a mean along $z$-direction, that is, around \(|0\rangle\). Then the variance can be determined by using 
\begin{equation}
    \sigma^2= \frac{\int \int \theta^2f(\theta,\phi)\sin{\theta} d\theta d\phi}{\int \int f(\theta,\phi)\sin{\theta}d\theta d\phi}, \label{eq6}
\end{equation}
where $\sigma$ represents the corresponding standard deviation. If we change the mean direction from $\ket{0}$ to $\ket{+}$, by keeping the other parameters of the disorder fixed, the corresponding value of the variance, $\sigma^2$, will remain unchanged. We use the standard deviation, $\sigma$, to measure the strength of the disorder that is present in the system.


\subsection{Haar-uniform distribution with a finite cut-off}
\label{Sub-A}
For this disorder distribution, 
the states, $\ket{\xi(\theta,\phi)}$, are uniformly and symmetrically distributed on the Bloch sphere, around $\ket{+}$. Thus we choose the angles $(\theta,\phi)$ Haar-uniformly within the parameter range $\theta\in[0,\pi]$ and $\phi\in[0,2\pi)$, and keep a required cut-off. 
The cut-off is dictated by physical considerations of the system under study, and is mathematically effected by 
defining a circular boundary outside of which the probability density function is set to zero. To realize the boundary, we imagine a plane parallel to the $y$-$z$ plane. That plane will cut the Bloch sphere, creating a circular boundary on the surface of the sphere. We will select only those points from the Haar-uniform generation which lie on the same side of the plane as the point \((1,0,0)\). 
Let $d$ be the angle - in radians - that the Bloch vector for $\ket{+}$ makes with the Bloch vector for an arbitrary state on the circular boundary.
This angle, $d$, can parameterize the range of the disorder. 


The variance of this  distribution is found using Eq.~\eqref{eq6}, and is given by
\begin{equation}
    \sigma^2 = \frac{\int_0^d \theta^2\sin\theta \;d\theta}{\int_0^d \sin\theta\;d\theta} = \frac{2d\sin d + 2\cos d - d^2\cos d - 2}{1 - \cos d}. \label{eq5}
\end{equation}
As expected, \(\sigma \to 0\) for \(d \to 0\).
The variance  increases with increasing $d$ and for $d = \pi/2$ and $\pi$, i.e. for disorder over the right half-sphere and the whole sphere, respectively, the standard deviations are $\sigma = 1.07$ and $1.71$.
All numbers are taken correct to three significant figures.

To analyze the situation in a numerical simulation, we consider a circle on the surface of the sphere, around the $z$-axis (instead of the \(x\)-axis). The circle is parallel to the $x$-$y$ plane and have the same area as the boundary of the distribution within which we want to generate the points. The boundary is specified using the angle $d$. 
Then we rotate each point, around the $y$-axis, to turn it into a distribution around the $x$-axis. The variance of such distribution can be calculated using Eq.~\eqref{eq5}.

\begin{figure*}
\centering
  \includegraphics[width=.45\linewidth]{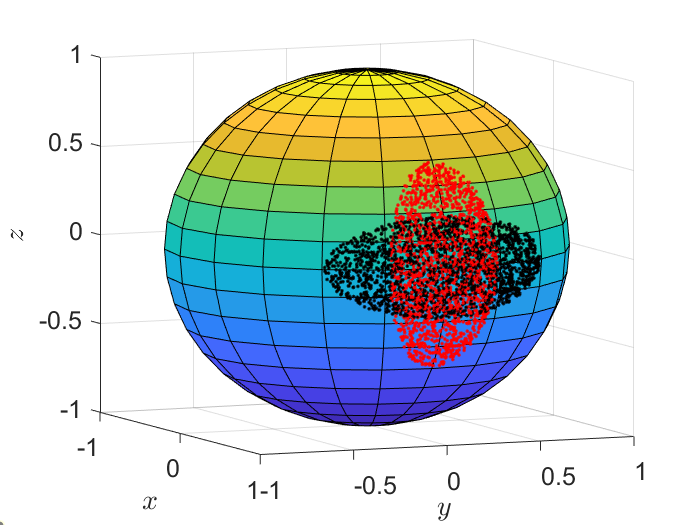}
  \label{fig:sub1}
  \includegraphics[width=.40\linewidth]{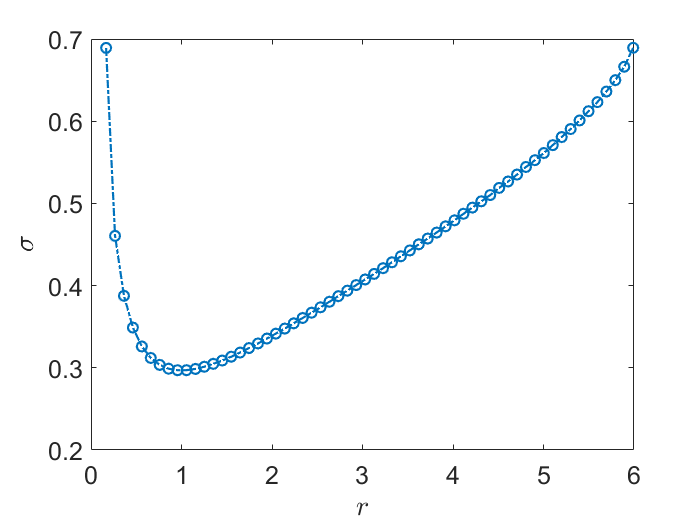}
\caption{Squeezed distribution and its standard deviation. On the left panel, we present two instances of the squeezed distribution on the surface of the Bloch sphere. The quantities describing the nature of the distribution are taken to be $D=0.524$, and $r=2$ (black points, horizontally-elongated patch) and $1/2$ (red points, vertically-elongated patch). 
On the right panel, the standard deviation is plotted along the vertical axis, against the squeezing parameter $r$ is on horizontal axis, for the squeezing distributions for which \(D = 0.524\). $\sigma$ attains the same value at a specific $r$ and its inverse $1/r$, and secures the minimum at $r = 1$, that corresponds to the special case of Haar-uniform distribution with a circular boundary, described using the parameter $d = \sin^{-1}\sqrt{D/\pi}$.
All quantities used are dimensionless.
}
\label{fig:squeezed_figure_and_sd}
\end{figure*}

\begin{figure*}
\centering
  \includegraphics[width=.45\linewidth]{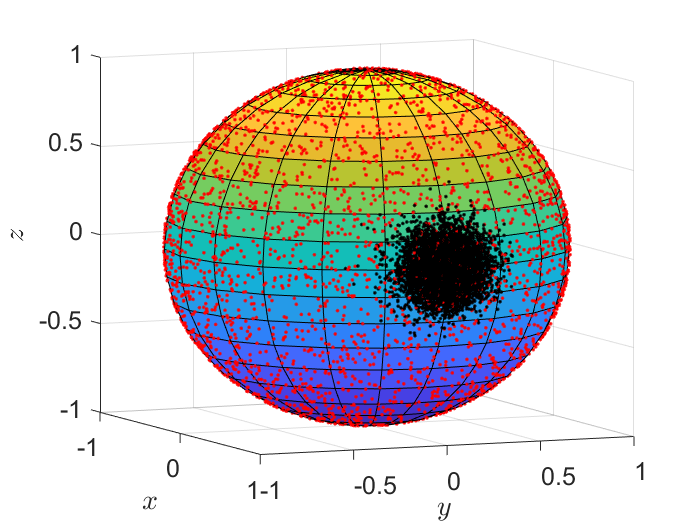}
  \label{fig:sub1-natun}
  \includegraphics[width=.40\linewidth]{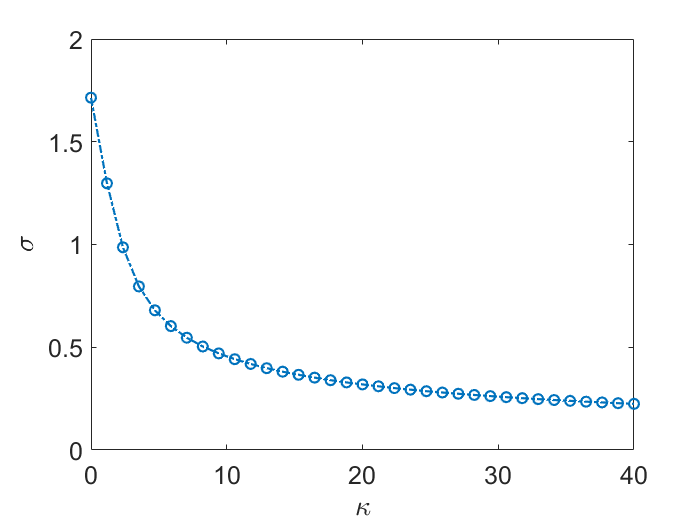}
\caption{Von Mises-Fisher distribution on a sphere and the nature of its standard deviation. 
On the left panel, randomly-chosen points are depicted on the Bloch sphere, from the von Mises-Fisher distribution, for two values of \(\kappa\), \(viz.\) $\kappa=70$ (black points, concentrated patch near \(|+\rangle\)) and $\kappa=0$ (red points, scattered over the whole surface of the sphere).
The mean of the distribution is specified to be along $x$ direction. 
The right panel depicts the behavior of the standard deviation with respect to the concentration parameter. All quantities used are dimensionless.
}
\label{fig:vonMises_figure_and_sd}
\end{figure*}

\subsection{Squeezed distribution}
\label{Sub-B}
In this case, the same distribution is considered as  in the preceding one, except that the current one is not symmetric about the mean, $\ket{+}$, i.e. the boundary here is not circular. The boundary can be squeezed along the $y$ or the $z$ axis. 
To obtain the squeezed boundary, we consider an ellipse on the $y$-$z$ plane (\(x=0\)). 
Let $a$ and $b$ denote the lengths of the semi-axes of the ellipse along the $y$ and $z$ axes respectively. $a=b=1$ represents the great circle on the Bloch sphere on the $y$-$z$ plane. We consider a cylinder parallel to the \(x\)-axis and having the ellipse as its cross-section. The cut-off (boundary) for the current distribution is defined as the curve obtained by the intersection of this cylinder and the Bloch sphere. 
We refer to the distribution as the ``squeezed" distribution. The area, \(D = \pi a b\), of the ellipse can characterize the spread of the distribution. However, even for a given \(D\), the variance, as given by Eq.~(\ref{eq6}), can vary.
The ratio $r = a/b$ encapsulates the degree of squeezing. The case when \(r\) is unity leads us back to the situation considered in the preceding subsection.

Given an area $D$, the corresponding range of allowed values of $r$ is obtained as follows:
\begin{enumerate}
    \item If $r \geq 1$, $a_{max} = 1$ and $b_{min} = D/\pi$;\hspace{2mm} $\therefore r_{max} = a_{max}/b_{min} = \pi/D$;\\
    \item If $r \leq 1$, $b_{max} = 1$ and $a_{min} = D/\pi$;\hspace{2mm} $\therefore r_{min} = a_{min}/b_{max} = D/\pi$.
\end{enumerate}
Here, $a_{max/min}$ and $b_{max/min}$ denote the maximum/minimum possible value of $a$ and $b$ respectively.
Hence the range of $r$ is $\left[ \frac{D}{\pi}, \frac{\pi}{D}\right]$. $r = 1$ is a special distribution which was discussed in Sec~\ref{Sub-A}. The angle $d$, defined for the $r=1$ case, is related to $D$ as $d = \sin^{-1}\sqrt{D/\pi}$.

We illustrate the distribution in 
Fig.~\ref{fig:squeezed_figure_and_sd}. In the left panel, we present a scatter diagram on the Bloch sphere of randomly generated points from the distribution for two values of $r$, $viz.$ $r=2$ and $r=1/2$, and a fixed value of $D$ \(viz.\) $D=0.524$. 
In the right panel of the same figure, we plot $\sigma$ as a function of $r$ for $D=0.524$.



\subsection{Von Mises-Fisher distribution}
\label{Sub-C}
The final disorder distribution on the Bloch sphere that we wish to consider is the ``spherical normal" one, also called the von Mises-Fisher distribution. The von Mises-Fisher distribution is an analogue of the usual normal distribution, but 
on the surface of a $(P-1)$-dimensional unit sphere in $\mathbb{R}^P$~\cite{vMF}. 
The probability density function  is given by
\begin{equation*}\label{von-Mises-dist}
    f_P(\pmb{x}; \mu,\kappa) = C_P(\kappa)\exp(\kappa\pmb{\mu}^T \pmb{x}),
\end{equation*}
where $\pmb{x}$ represents the coordinates of a randomly chosen point on the unit sphere. The parameters $\pmb{\mu}$ and $\kappa$ denote the direction of mean and the concentration of points on the sphere, respectively, satisfying $\kappa \geq 0$ and $\left|\left|\pmb{\mu}\right|\right| = 1$. 
The normalization constant $C_P(\kappa)$ is $\frac{\kappa^{P/2 - 1}}{(2\pi)^{P/2}I_{P/2 - 1}(\kappa)}$ with $I_{P/2 - 1}(\kappa)$ being the modified Bessel function of first kind at order $(P/2 - 1)$.  The distribution is unimodal for $\kappa > 0$, and is uniform on the surface of the sphere, for $\kappa = 0$. As $\kappa$ increases the distribution becomes more and more concentrated around its mean.

We want to generate points, $(1,\theta,\phi)$, on the Bloch sphere. Thus the distribution of our interest is the von Mises-Fisher distribution for $P=3$, which is given by
\begin{equation*}
    f_3(\pmb{x}; \mu,\kappa) = \frac{\kappa}{4\pi\sinh\kappa} \exp(\kappa\pmb{\mu}^T \pmb{x}).
\end{equation*}
We take the direction of the mean axis to be $\mu=(1,\pi/2,0)$. The generation of points is discussed in Appendix~B. 
We find that \(\sigma \to 0\) for \(\kappa \to \infty\), and \(\sigma \to 1.71\) for \(\kappa \to 0\). Note that \(\kappa \to 0\) leads us to the Haar-uniform distribution over the entire Bloch sphere, which is obtained when \(d=\pi\) in Sec.~\ref{Sub-A}.


We illustrate the distribution in Fig.~\ref{fig:vonMises_figure_and_sd}, where we depict randomly-chosen points from the von Mises-Fisher distribution, for $\kappa=0$ and $\kappa=70$. The mean direction of these points are restricted to be along $x$-axis. These points are plotted on the surface of the Bloch sphere in the left panel of Fig.~\ref{fig:vonMises_figure_and_sd}.
The standard deviations (Eq.~\eqref{eq6}) for different \(\kappa\) are plotted 
in the right panel of Fig.~\ref{fig:vonMises_figure_and_sd}.

\section{Behavior of algorithm in response to disorder }
\label{sec6}
In this section, we consider each of the three  disorder distributions considered in the preceding section separately, and explore their impact  on the QPE process. Precisely, we determine the disorder-averaged probability, $q_j$, and examine its nature with varying strength of the corresponding disorder.
Let us spend a few sentences here for describing the disorder averaging in the glassy disorder case that we are considering. 
For a given disorder distribution with a fixed set of values for its function  parameters,
we independently select $m$ pairs of angles $(\theta_i,\phi_i)$ within the allowed range and evaluate $p_j'$ using Eq.~\eqref{prob_j}. This process is repeated multiple times and we take the average of these $p'_j$s to obtain the disorder-averaged probability, $q_j$. 
The order of first calculating a system characteristic and only subsequently calculating the average over the disorder configurations, is exactly what is physically relevant for glassy systems, as typical observation times are orders of magnitude lower than equilibration times of the disorder. 
To check for convergence, we need to determine $q_j$ for a larger set of $p_j's$, and compare the values. We will check for convergence up to three significant figures.


\subsection{Uniform disorder}
In this part, we will determine the effect of uniform disorder on the QPEA.
To be precise, we will consider a noisy Hadamard gate whose output corresponding to the input $\ket{0}$, instead of being $\ket{+}$, is $|\xi(\theta,\phi)\rangle$, where the angles $(\theta,\phi)$ are such that the points  $(1,\theta,\phi)$ are Haar uniformly distributed on the surface of the Bloch sphere, with a circular cut-off. As we have mentioned previously, the amount of disorder will be quantified using the standard deviation, $\sigma$, of the disorder distribution. In the following part (Sec.~\ref{sub1}), we will consider two specific values of $\sigma$ and determine the corresponding $q_j$. In the next portion, (Sec.~\ref{sohag-chand}), the noisy QPEA will be explored 
for a broader set of values of $\sigma$.
\subsubsection{Two special cases}
 \label{sub1}
The two specific cases that we investigate separately here are respectively those for which the 
points $(1,\theta,\phi)$  are distributed over the whole sphere, i.e. $d=\pi$ ($\sigma = 1.71$), and for which  $d=\pi/2$ ($\sigma = 1.07$) i.e. the points are distributed over one-half of the sphere with the noiseless case being placed symmetrically in the middle of that half-sphere. 

For $m = 1$, the $j$ of Eq.~\eqref{prob_j} can take only two distinct values, $viz.$ 0 and 1. Thus, we can evaluate the phase up to only one binary digit, i.e. the phase will be known to be either $p=0$ or $p=1/2$. In case of disorder over the whole sphere, the corresponding disorder-averaged probabilities are $q_0=q_1 = 1/2$.


Similarly, for $m = 5$, the phase can be estimated up to five binary digits, i.e. $p = j/2^5$, and the corresponding values of $j$ are $0,1,\ldots,31$. In this case, $q_j = 
1/2^5$, for all $j$.

For an arbitrary number of auxiliary qubits, say $m$, the value of the disorder-averaged probability is $q_j = 1/2^m$, for $j = 0,1,...,2^m-1$. 
For \(d=\pi\), $q_j$ is independent of the error $\delta=p-1/2^m$ as well as $j$, which means that whatever be the phase, all the outcomes are equally probable. Thus the QPEA cannot work in presence of uniform disorder over the entire Bloch sphere.

In case of disorder over half of the sphere, the disorder-averaged probability of a certain outcome $\frac{j}{2^m}$, for arbitrary number, \(m\), of auxiliary qubits  is given by
\begin{equation*}
    q_j = \frac{1}{2^{2m}}\prod_{i=1}^{m}\left[2 + \cos 2^i\pi\left(p - \frac{j}{2^m}\right) \right],
\end{equation*}
where $j = 0,1,..., 2^m - 1$. Let the actual phase be $p = \frac{j}{2^m} + \delta$, where $\delta \leq 2^{-(m+1)}$. Then, for example if we take $m = 5$ and $\delta = 1/2^{10}$, $q_j$ is approximately equal to 0.237. Compare with 
Fig.~\ref{fig:uniform_plot}. Note that for disorder over the half sphere, $\sigma = 1.07$.


\subsubsection{Effect of uniform disorder of arbitrary strength}
\label{sohag-chand}

\begin{figure}[h!]
    \centering
    \includegraphics[scale = 0.40]{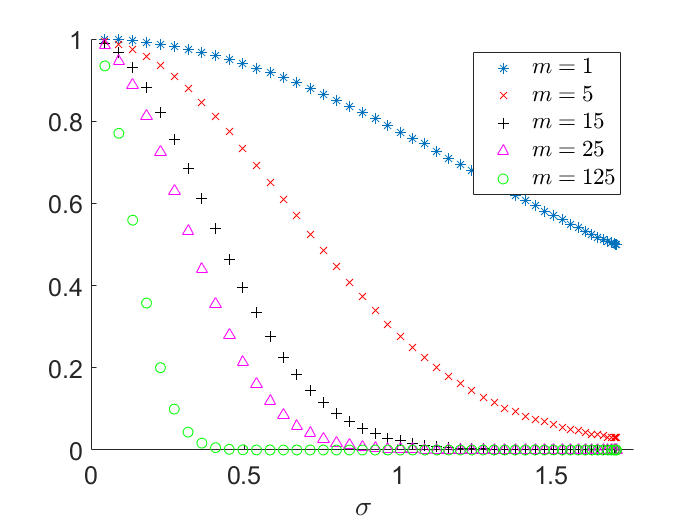}
    \caption{Response of success probability of QPEA to uniform disorder. We plot here the disorder-averaged probability in presence of uniform disorder with a finite cut-off 
    with respect to the standard deviation $\sigma$ of the distribution
    for different numbers, \(m\),  of auxiliary qubits utilized in the algorithm. 
    The horizontal axis represents $\sigma$ and the corresponding disorder-averaged probabilities are plotted along the vertical axis. Both axes are dimensionless.}
    \label{fig:uniform_plot}
\end{figure}
\begin{figure*}[t]
    \centering
    \includegraphics[scale=0.54]{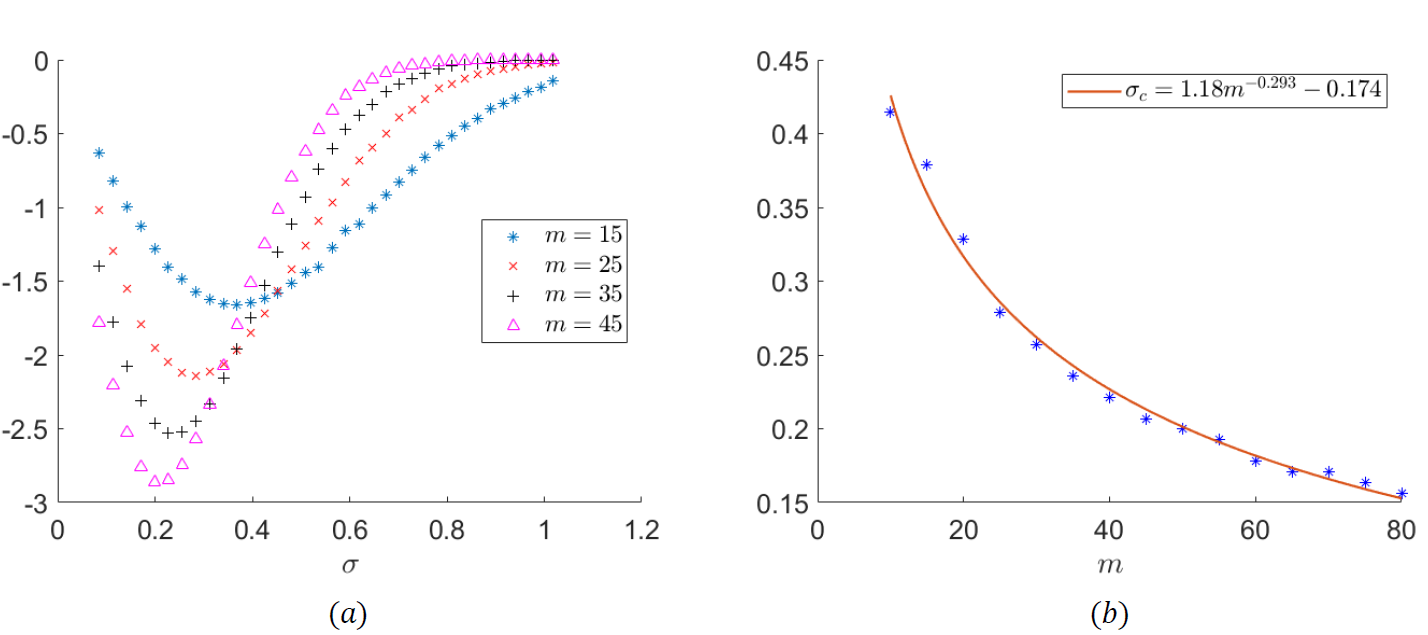}
    \caption{Scaling of weak-to-strong disorder changeover with number of auxiliary qubits in the QPEA. 
    (a) All considerations in this panel are the same as in Fig.~\ref{fig:uniform_plot}, except that the vertical axis here represents the derivatives of the disorder-averaged probabilities.
    Also a different set of \(m\) are exemplified, as mentioned in the legend. 
    (b) In this panel, the vertical axis represents $\sigma_c$, the minima of the curves in the left panel. This is the point where the corresponding curve in Fig.~\ref{fig:uniform_plot} changes its curvature. The horizontal axis represents $m$. Both the axes are dimensionless. The blue stars represent the numerically obtained data, whereas Eq.~\eqref{eq2} is plotted using the red curve.}
    \label{fig:uniform_der_plot}
\end{figure*}

\begin{figure*}
    \centering
    \includegraphics[scale = 0.54]{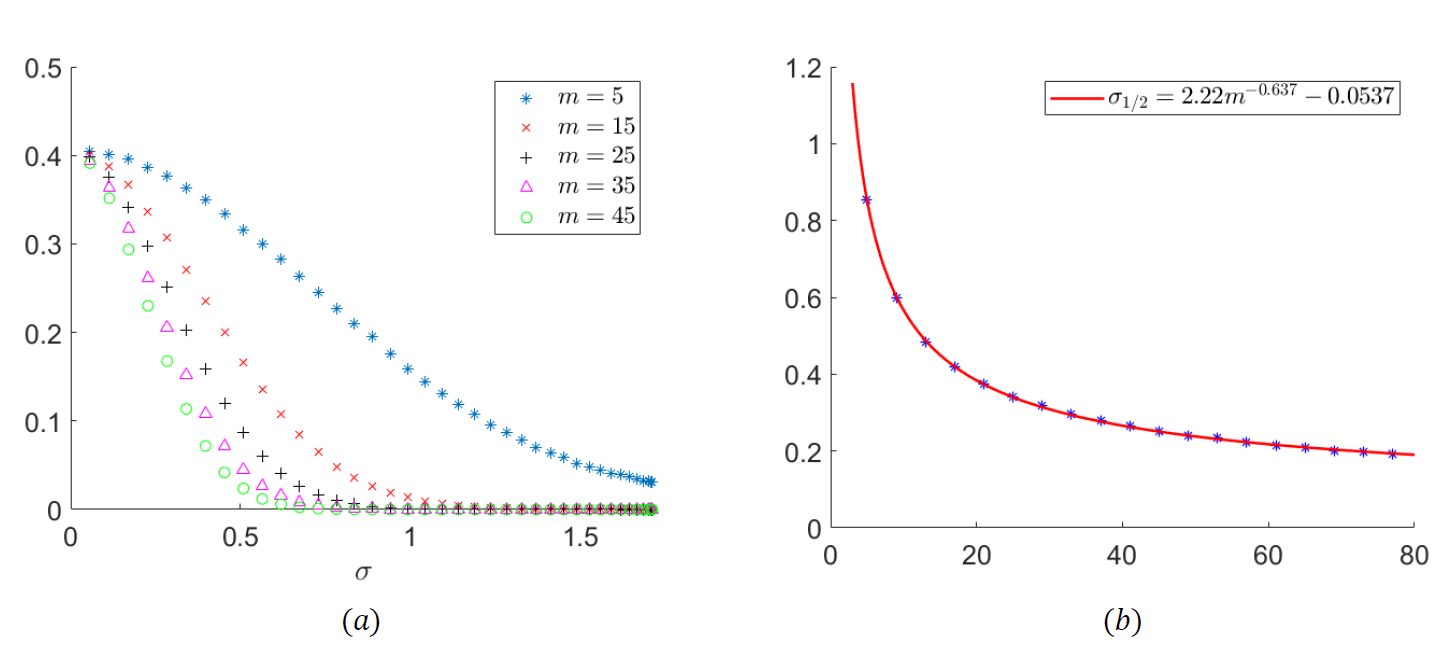}
    \caption{Speed of decay of disorder-averaged success probability with number of auxiliary qubits. 
    (a) The considerations in this panel are the same as in Fig.~\ref{fig:uniform_plot}, except that the vertical axis here denotes $q_{min}$, and a different set of \(m\) are considered here, as mentioned in the legend. 
    (b) In this panel, the vertical axis denotes $\sigma_{1/2}$, the disorder at which the disorder-averaged probability, $q_{min}$, reduces to half of  $p_{min}$,  the minimum probability in the noiseless situation. The horizontal axis represents $m$. The data points are plotted using blue stars. These points are then fitted with the curve given in Eq.~\eqref{eq3}, shown in the figure using a red line.}
    \label{fig:prob_all_vs_d}
\end{figure*}

Let us now proceed to the more general case, where the distribution of the disorder is still Haar-uniform but the strength of disorder, $\sigma$, is varied more flexibly. The parameter, $d$, fixes the range of the distributed angles, $(\theta,\phi)$. 
In the computation of $q_j$, we choose $\delta$ according to the following list:
\begin{center}
\begin{tabular}{ |c|c|c|c|c|c|c| } 
\hline
$m$ & 5 & 15 & 25 & 35 & 45 & 125 \\
\hline
$\log_2\delta$ & {-10} & {-20} & {-30} & {-40} & {-50} & {-130}\\  
\hline
\end{tabular}
\end{center}
We plot the results in Fig.~\ref{fig:uniform_plot}.
The figure shows the dependence of $q_j$ on $\sigma$, for different number of auxiliary qubits, $m$. It can be observed from the figure that when $\sigma$ is small, $q_j$ decreases slowly with increasing $\sigma$. When the disorder is increased further, suddenly the probability starts to decrease drastically. The threshold value of $\sigma$, at which the change from the slow decrease to fast decrease takes place, depends on $m$. Higher the value of $m$, smaller is the corresponding threshold $\sigma$. For high disorder strengths, the probability becomes almost constant. This 
saturated value, say $p_{sat}$, also depends on $m$. For sufficiently large $m$, $p_{sat}\approx 0$. E.g. for $m \geq 15$, $q_j\approx 0$ for $\sigma > 1.07$ (i.e. for $d > \pi/2)$, i.e. when the output states, $\ket{\xi(\theta,\phi}$, are distributed over or beyond the half-sphere).

\begin{figure}[t]
    \centering
    \includegraphics[scale=0.45]{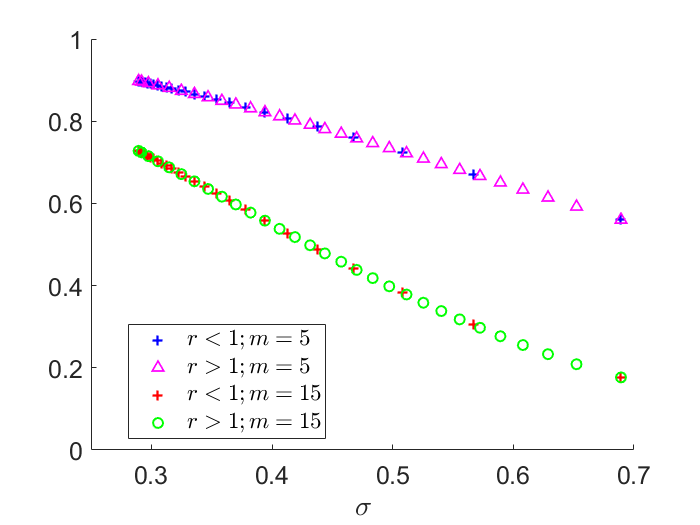}
    \caption{Response of QPEA to squeezed disorder. 
    We plot $q_j$ along the vertical axis, with respect to the standard deviation, $\sigma$, of the distribution of points, represented along the horizontal axis. Blue (red) $+$ points denote the case when $r < 1$, i.e. squeezing along $z$-axis and magenta (green) $\Delta$ ($\circ$) points represent the $r > 1$ case, i.e. squeezing along $y$-axis, where the number of available auxiliary qubits are fixed to  $m=5$ ($m=15$). The plots correspond to the case when area of the elliptical projection of distributed points on the $y$-$z$ plane, is $D=1/2$. All quantities used are dimensionless.
    }
    \label{fig:ellip_m15_area05_fullr}
\end{figure}

For each value of $m$, the corresponding curve, in Fig.~\ref{fig:uniform_plot}, changes from concave to convex at a particular $\sigma$.
To determine the inflection point, $\sigma_c$, we present the derivatives of the disorder-averaged probabilities with respect to $\sigma$ in Fig.~\ref{fig:uniform_der_plot}(a), for different values of $m$. The 
minimum of each of the curves in Fig.~\ref{fig:uniform_der_plot} (a), indicates the point, $\sigma_c$. There is clearly a dependence of $\sigma_c$ on $m$. For higher value of $m$, $\sigma_c$ gets closer to zero. For a given \(m\), we can define two regions on the \(\sigma\) axis: ``weak" disorder, when the strength of disorder as defined by $\sigma$ is much smaller than $\sigma_c$, and ``high" disorder, when $\sigma$ is significantly larger than $\sigma_c$. In Fig.~\ref{fig:uniform_der_plot}(b), we plot $\sigma_c$ versus $m$, which indicates a log-log dependence. 
We can fit the curve using the following function:
\begin{equation}\label{eq2}
    \ln(\sigma_c - \alpha) = \ln\beta + \gamma \ln m,
\end{equation}
where the fitting parameters are given by $\alpha=-0.174\pm 0.314$, $\beta=1.18 \pm 0.0790$, and $\gamma=-0.293\pm 0.209$. The curve is fitted using the least squares method (see Appendix~C for more details). The least square error obtained in this fitting is $6.20\times 10^{-5}$. Numbers written after the $\pm$ sign denote the 95\% confidence interval of the corresponding parameter's value.

As stated in section \ref{sec2}, in the disorder-free case,
$p_{min}\sim 0.4$ for large $m$.
Fig. \ref{fig:prob_all_vs_d}(a) depicts how the disorder-averaged value of $p'_{min}$, i.e. $q_{min}$, varies with the amount of disorder, $\sigma$. In the noiseless situation, i.e. for $\sigma = 0$, $q_{min}$ take values close to $0.4$ for all values of $m$. It starts to decrease with increasing $\sigma$. This degradation becomes faster for higher values of $m$. The pattern of the curves in Fig.~\ref{fig:prob_all_vs_d}(a) are similar to those in Fig.~\ref{fig:uniform_plot}.
We see that for a large number of auxiliary qubits, the noise from each Hadamard gate adds up, greatly affecting the qubits, and thus the collective impact of the noise dominates over the benefit obtained by using large $m$.

Let us now compute the disorder strength at which the disordered-averaged probability, $q_{min}$, reduces to half of its original value, i.e. the value in the noiseless case, $p_{min}$. We denote this disorder strength by $\sigma_{1/2}$. 
The dependence of $\sigma_{1/2}$ on $m$ is plotted in Fig.~\ref{fig:prob_all_vs_d}(b). This gives an impression about the speed with which the probability falls off with increasing number of auxiliary qubits. This curve also has a log-log dependence (like in Fig.~\ref{fig:uniform_der_plot}(b)). The following function is used to fit the curve:
\begin{equation}\label{eq3}
    \ln(\sigma_{1/2}-\alpha) = \ln\beta +\gamma \ln m,
\end{equation}
where $\alpha=0.0537\pm 0.0.00905$, $\beta=2.22 \pm 0.0450$, and $\gamma=-0.637\pm 0.0167$. The least-square error for the fitting is found to be $4.25\times 10^{-6}$.


\subsection{Squeezed disorder}
\label{5-SubB}

Let us now discuss the impact of disorder following squeezed distribution on the probability of obtaining the correct outcome in the QPEA. We plot $q_j$ as a function of $\sigma$, in
Fig.~\ref{fig:ellip_m15_area05_fullr}, for two fixed numbers of auxiliary qubits, \(viz.\) $m=5$ and $m=15$, and for a fixed area, $D=1/2$, of the elliptical projection. It can be noticed from the figure that the value of the disorder-averaged probability is same for a given $\sigma$ irrespective of the squeezing direction, i.e. whether $r < 1$ or $r > 1$. The probability does  not  distinguish between whether the direction of squeezing is along the $y$- or the $z$-axis, even though there is a bias in the choice of the initial state of auxiliary qubits (which are taken as $\ket{0}$), and the control unitary gate - the $\ket{k}$ in Eq.~(\ref{je-katha-tomake-bolbo}) are tensor products of eigenvectors of the Pauli-$z$ operator.

\begin{figure}[h!]
    \centering
    \includegraphics[scale = 0.43]{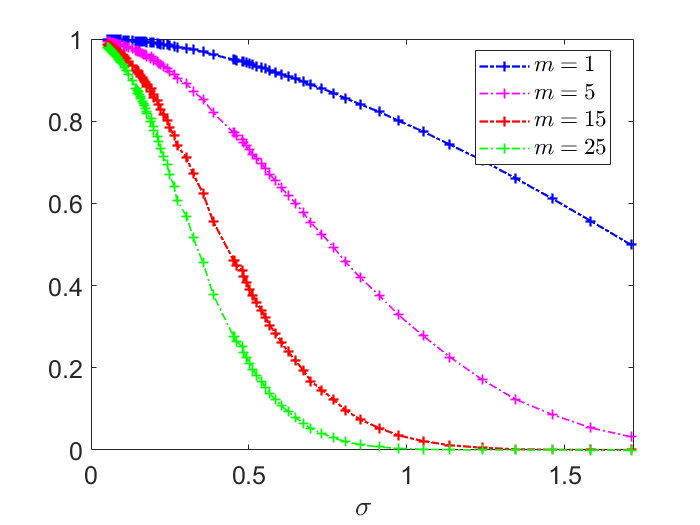}
    \caption{Response of QPEA to spherical normal disorder distribution. 
    We plot the disorder-averaged probability $q_j$ along the vertical axis with respect to the standard deviation, $\sigma$, of the von Mises-Fisher distribution represented along the horizontal axis, for different numbers of auxiliary qubits used in the QPEA. All quantities used are dimensionless.}
    \label{fig:von-Mises}
\end{figure}

\begin{figure}[h!]
    \centering
    \includegraphics[scale = 0.43]{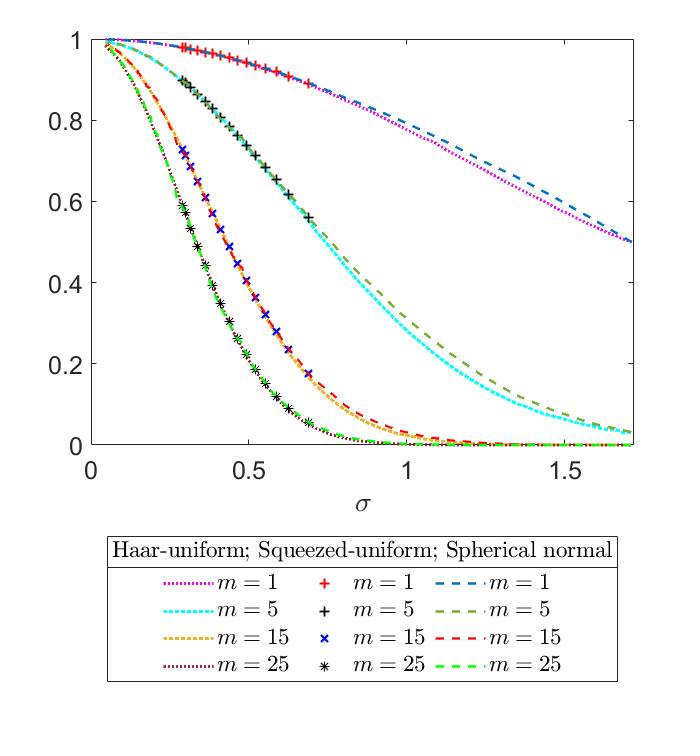}
    \caption{Resemblance of disorder-averaged probabilities for Haar-uniform, squeezed-uniform and spherical normal disorders. We present the values of $q_j$ (along the vertical axis) with respect to the standard deviation  $\sigma$ (along the horizontal axis) of the corresponding distribution, for different values of $m$. We use different coloured dotted lines for Haar-uniform, point types for squeezed uniform, and dash-dotted lines for spherical normal distribution, as mentioned in the legend. The axes are dimensionless.}
    \label{fig:comparison_plot}
\end{figure}

\subsection{Spherical normal disorder}

We now consider the case when the disorder follows the von Mises-Fisher distribution. Fig.~\ref{fig:von-Mises} shows the variation of  disorder-averaged probability with respect to  standard deviation $\sigma$ of the spherical normal disorder distribution.
Qualitatively, $q_j$ gets affected by spherical normal disorder in the same way as it is in case of uniform disorder. Specifically, $q_j$ takes values near unity, for all $m$ in the noiseless situation, and decreases monotonically with increasing $\sigma$. And there is a change in curvature from concave to convex at a disorder strength, whose value depends on \(m\). The disorder also affects more  when the number of auxiliary qubits is large.

\subsection{Comparison between the responses to different disorder distributions}
To analyze the similarities between the effects of different disorder distributions on the QPEA, we plot in Fig.~\ref{fig:comparison_plot},  the disorder-averaged probabilities, $q_j$, as  functions of the standard deviations, $\sigma$, of the three types of incorporated disorders (studied in the three preceding subsections). The coordinates $(\sigma,q_j)$ corresponding to uniform, squeezed, and spherical normal disorders are plotted using dotted lines, symbol points, and dashed lines respectively. 
In accordance to the Corollary in Sec.~\ref{sec4}, the plots fall on each other for relatively large \(m\). And so the three curves for \(m=25\) are almost indistinguishable from each other. However, what is not covered in the Corollary is that for all \(m\), the disorder-averaged success probabilities match, whenever the strength of the disorder is very strong or very weak. There is only an intermediate range of \(\sigma\), for which the probabilities differ. For example, $q_j$ for uniform disorder deviates from that for spherical normal disorder, within the range $\sigma \in (0.8, 1.6)$ approximately. We therefore find that an experimental differentiation between the disorder that is active in the Hadamard unitary element in the QPEA circuit can be efficiently performed only when the number of qubits is rather small, and when the disorder strength is in the moderate range.


\section{conclusion}
\label{sec7}
The quantum phase estimation algorithm lies  at the heart of several applications in quantum information and computation.
In realistic scenarios, both environmental noise and noise due to imperfections  within the circuit affect the performance of the algorithm in estimating the correct phase value up to a desired precision. We investigated the effects of ``glassy disorder" originating from  faulty Hadamard gates used in  the quantum phase estimation circuit.

{ We analyzed - analytically and numerically - the disorder-averaged success probability of measuring the correct phase value, and its dependence on the strength of disorder. We proved that for a sufficiently large number of auxiliary qubits, the disorder-averaged probability does not depend on the type of disorder distribution, provided the distribution has certain simple properties.
}
As examples, we studied three disorder distributions, \textit{viz.} Haar-uniform with a finite circular cut-off, Haar-uniform with an elliptical (squeezed) cut-off, and von Mises-Fisher disorder. We found that the depreciation of the success probability from its ideal value depends on  the number of auxiliary qubits used in the algorithm. We also saw that the disorder-averaged probability changes its curvature from concave to convex as a function of the disorder strength, indicating a marked increase in the speed of depreciation after passing this point of inflection. In case of uniform disorder, we estimated a log-log dependence between the inflection point and the number of performing auxiliary qubits. 

{The responses to the different types of disorder distributions in the success probability are qualitatively similar, and often even quantitatively. We analytically proved the quantitative similarity for a large number of auxiliary qubits.  For a moderate or small number of auxiliaries, restricting to three types of disorder distributions, we numerically observed a qualitative similarity for all disorder strengths, and even a quantitative similarity for weak  and strong disorders (as defined by a non-moderate standard deviation of the distribution). Experimental detection of the disorder distribution that is active in the circuit can therefore be efficiently possible only when the number of auxiliary qubits is not high and when the disorder strength is moderate.}
%
Our work is potentially a step toward creating a dictionary for an experimenter of her expectation of the capability of the quantum phase estimation process in a realistic, noisy scenario.

\begin{acknowledgments}
S.M. acknowledges support from the KVPY program. We acknowledge partial support from the Department of Science and Technology, Government of India through the QuEST grant (grant no. DST/ICPS/QUST/Theme-3/2019/120). 
\end{acknowledgments}

\section*{Appendix A}
\label{appendix-A}
In the quantum phase estimation algorithm, the initial state of the auxiliary qubits is considered to be $\ket{\zeta_0}=\ket{0}^{\otimes m}$. After  application of the noisy $H$ gates, the state of the system of the $m$ auxiliary qubits becomes
\begin{multline*}
    \ket{\zeta'_1}=H_m\ket{0}...H_1\ket{0}\\
    = \left(\cos\frac{\theta_m}{2}\ket{0} + e^{i\phi_m}\sin\frac{\theta_m}{2}\ket{1} \right)\otimes\\
    \cdots \otimes\left(\cos\frac{\theta_1}{2}\ket{0} +e^{i\phi_1}\sin\frac{\theta_1}{2}\ket{1} \right),
\end{multline*}
where $\theta_i$ and $\phi_i$ describe the independent local disorder, present in the noisy Hadamard gate $H_i$. Operation of the controlled-U gate on $\ket{\zeta'_1}$, produces the state
\begin{eqnarray*}
    \ket{\zeta'_2}=\Lambda_m(U)\ket{\zeta'_1}=  \left(\cos\frac{\theta_m}{2}\ket{0} + e^{2^m\pi ip} e^{i\phi_m}\sin\frac{\theta_m}{2}\ket{1} \right)\otimes\\
    \cdots \otimes\left(\cos\frac{\theta_1}{2}\ket{0} + e^{2\pi ip}e^{i\phi_1}\sin\frac{\theta_1}{2}\ket{1} \right).
\end{eqnarray*}
This can be written in terms of $\ket{k}$
as
\begin{multline}\label{lambdaU_state_k}
     \ket{\zeta'_2}=\cos\frac{\theta_m}{2}\cdots \cos\frac{\theta_2}{2} \cos\frac{\theta_1}{2}\ket{0}\\
     + e^{2\pi ip + i\phi_1} \cos\frac{\theta_m}{2}\cdots \cos\frac{\theta_2}{2} \sin\frac{\theta_1}{2}\ket{1}\\ + \cdots + e^{(2^m-1)2\pi ip + i(\phi_1 + \cdots+\phi_m)} \sin\frac{\theta_m}{2}\cdots \sin\frac{\theta_1}{2} \ket{2^m - 1}.
\end{multline}
Set \(\ket{\zeta'_2}
\sum_{k=0}^{2^m-1}y_k\ket{k}\). The application of $\text{QFT}^\dagger_{2^m}$ on $\sum_{k=0}^{2^m-1}y_k\ket{k}$ will produce the state $\ket{\zeta'_3}=\text{QFT}^\dagger_{2^m}\ket{\zeta'_2}=\text{QFT}^\dagger_{2^m}\sum_{k=0}^{2^m-1}y_k\ket{k} =\sum_{j=0}^{2^m-1}x_j\ket{j}$, where the coefficients, $x_j$, are given by
\begin{align*}
    x_j &= \frac{1}{\sqrt{2^m}}\sum_{k=0}^{2^m -1} y_k e^{-\frac{2\pi ijk}{2^m}} \\
    &= \frac{1}{\sqrt{2^m}} \prod_{l=1}^{m}\left[ \cos\frac{\theta_l}{2} + e^{2^l\pi i(p - \frac{j}{2^m}) + i\phi_l}\sin\frac{\theta_l}{2} \right].
 \end{align*}
On measurement of the auxiliary states in the computational basis $\{\ket{j} \}_{j=0}^{2^m-1}$, we obtain the state $\ket{j}$ with probability
\begin{equation*}
    p_j' = \left|{x_j}\right|^2 = \frac{1}{2^m}\prod_{l = 1}^{m}\left[ 1 + \sin\theta_l \cos\left[2^l \pi\left(p - \frac{j}{2^m}\right) + \phi_l\right] \right].
\end{equation*}

\section*{Appendix B}
\label{appendix-B}
Generating random variates from the spherical normal distribution: We begin by generating a distribution of points around the $z$-axis on Bloch sphere. Hence $\pmb{\mu} = (0,0,1)^T$ and $\pmb{x} = (\sin\theta\cos\phi,\sin\theta\sin\phi,\cos\theta)^T$, in Cartesian coordinates, which give $\pmb{\mu}^T \pmb{x} = \cos\theta$.  
Here, $p = 3$. The normalization constant here is
\begin{equation*}
    C_3(\kappa) = \frac{\sqrt{\kappa}}{(\sqrt{2\pi})^{3} I_{1/2}(\kappa)} = \frac{\kappa}{4\pi\sinh\kappa},
\end{equation*}
and $I_{1/2} = \sqrt{\frac{2}{\pi \kappa}}\sinh\kappa$. Thus the required probability density function is 
\begin{equation*}
    f_3(\theta) = \frac{\kappa}{4\pi\sinh\kappa} \exp(\kappa\cos\theta).
\end{equation*}

Since we want to generate points that follow the above distribution, we first compute the cumulative distribution function,

\begin{equation*}
    F(\theta) = \int_0^\theta 2\pi f_3(\theta') \sin\theta' d\theta'    = \frac{e^\kappa - e^{\kappa\cos\theta}}{2\sinh\kappa}.
\end{equation*}
The range $\theta \in [0,\pi]$ confirms $F(\theta) \in [0,1]$. For a certain constant $A$, if $F(\theta) = A$, the inverse is given by
\begin{equation*}
    \theta = \cos^{-1} \left[ \frac{1}{\kappa} \ln(e^\kappa - 2A\sinh\kappa) \right] = F^{-1}(A).
\end{equation*}
Choosing $A$ from a uniform distribution within the range $[0,1]$, we can generate $\theta$ that follows the von Mises-Fisher distribution. All that remains is to select $\phi$ randomly from the uniform distribution over the range $[0,2\pi)$ to create a distribution of points around the $z$-axis. Then we rotate each point by operating the matrix
\begin{equation*}
    \textbf{A} = \begin{pmatrix}
         0 & 0 & 1\\
         0 & 1 & 0\\
        -1 & 0 & 0
    \end{pmatrix}
\end{equation*}
on $(\sin\theta\cos\phi,\sin\theta\sin\phi,\cos\theta)^T$ (in Cartesian coordinates), and turn it into a distribution of points around the $x$-axis.

\section*{Appendix C}
\label{appendix-c}
We used the \textit{least squares method} to find the best-fitting curve 
to the set of data points $\{((\sigma_\chi)_i, m_i)\}$ of size $N$, obtained from numerical calculations. The functional form of the fit was taken as 
\begin{equation*}
    \sigma_\chi(m|\alpha,\beta,\gamma) = \alpha + \beta m^\gamma,
\end{equation*}
where $\sigma_\chi$ is either $\sigma_c$ or $\sigma_{1/2}$. Taking natural logarithms on both sides, we can convert it to a linear form:
\begin{equation*}
    \ln(\sigma_\chi-\alpha) = \ln\beta + \gamma\ln m.
\end{equation*}
In all cases considered, \(\sigma_\chi - \alpha\) turned out to be positive in the relevant range.
Using the least squares method, we find the parameters $\alpha$, $\beta$, and $\gamma$ that minimize the mean square error
\begin{equation*}
    D = \frac{1}{N}\sum_i \left[ \ln (\sigma_\chi - \alpha) - \ln ((\sigma_\chi)_i - \alpha)\right]^2,
\end{equation*}
so that the least error is $MSE = min_{\alpha,\beta,\gamma} D$.

\section*{Appendix D}
\label{appendix-D}

To prove that the minimum probability to successfully gather information about the phase corresponds to $p - \frac{j}{2^m} = \delta = 2^{-(m+1)}$, we consider the case $\delta \neq 0$, where
\begin{equation*}
    p_j = 2^{-2m}\frac{1-\cos(2\pi2^m\delta)}{1-\cos(2\pi\delta)}.
\end{equation*}
The range of this function (of \(\delta\)) is $\delta \in [-2^{-(m+1)},2^{-(m+1)}]$. Moreover, it is an even function. So it is enough to prove that the function $f(\delta) = \frac{1-\cos(2\pi2^m\delta)}{1-\cos(2\pi\delta)}$ is a monotonically decreasing function in the range $\delta \in (0,2^{-(m+1)}]$, i.e., the derivative of $f(\delta)$ with respect to $\delta$ is negative (in that range). Thus we need to show that
\begin{equation}\label{last-eqn-maybe}
    \frac{\sin(2\pi\delta)}{\cos(2\pi\delta)-1} < 2^m \frac{\sin(2\pi2^m\delta)}{\cos(2\pi2^m\delta)-1}.
\end{equation}
However, the left and right hand sides of the inequality consist of a function of the form $g(x) = \frac{\sin x}{\cos x-1}$. The derivative of this function with respect to $x$ is $g'(x) = \frac{1}{1-\cos x} > 0$, for $x \in (0,\pi)$, i.e., $g(x)$ is a monotonically increasing function in that range. Hence we have $g(2\pi\delta) < g(2\pi2^m\delta)$, and so $g(2\pi\delta) < 2^m g(2\pi2^m\delta)$. This corresponds to Eq.~\eqref{last-eqn-maybe}, which proves that the minima of $p_j$ corresponds to $\delta = 2^{-(m+1)}$.

\end{document}